\documentclass[aps,prl,graphicx,9pt,twocolumn,superscriptaddress]{revtex4-2}
\usepackage[colorlinks,linkcolor=blue,urlcolor=blue,anchorcolor=blue,citecolor=blue]{hyperref}
\usepackage{amsmath}
\usepackage{graphicx}
\usepackage{dcolumn}
\usepackage{mathrsfs}
\usepackage{amssymb}
\usepackage{amsfonts,multirow}
\usepackage{bm,soul,lineno}
\usepackage{extarrows}

\usepackage{epstopdf}
\usepackage{amsmath}
\usepackage{graphicx}
\usepackage{dcolumn}
\usepackage{verbatim}
\usepackage{mathrsfs}
\usepackage{amssymb}
\usepackage{amsfonts}
\usepackage{bm}

\newif\ifistoreview

\begin{document}
	\draft
    \title{A hypersphere-like non-Abelian Yang monopole and its topological characterization}
	\author{Shou-Bang Yang}
            \affiliation{Fujian Key Laboratory of Quantum Information and Quantum
        		Optics, College of Physics and Information Engineering, Fuzhou University,
        		Fuzhou, Fujian, 350108, China}
        \author{Pei-Rong Han} 
            \affiliation{School of Physics and Mechanical and Electrical Engineering, Longyan University, Longyan, China}
	\author{Wen Ning}
            \affiliation{Fujian Key Laboratory of Quantum Information and Quantum
        		Optics, College of Physics and Information Engineering, Fuzhou University,
        		Fuzhou, Fujian, 350108, China}
	\author{Fan Wu}
            \affiliation{Fujian Key Laboratory of Quantum Information and Quantum
    		Optics, College of Physics and Information Engineering, Fuzhou    University,
    		Fuzhou, Fujian, 350108, China}
	\author{Zhen-Biao Yang} \email{zbyang@fzu.edu.cn}
            \affiliation{Fujian Key Laboratory of Quantum Information and Quantum
		Optics, College of Physics and Information Engineering, Fuzhou University,
		Fuzhou, Fujian, 350108, China}
	\author{Shi-Biao Zheng} \email{t96034@fzu.edu.cn}
            \affiliation{Fujian Key Laboratory of Quantum Information and Quantum
		Optics, College of Physics and Information Engineering, Fuzhou University,
		Fuzhou, Fujian, 350108, China}
	
	\vskip0.5cm
	
	\narrowtext
	
	\begin{abstract}
		Synthetic monopoles, which correspond to degeneracies of Hamiltonians, play a central role in understanding exotic topological phenomena. Dissipation-induced non-Herminicity (NH), extending the eigenspectra of Hamiltonians from the real to complex domain, largely enriches the topological physics associated with synthetic monopoles. 
        We here investigate exceptional points (EPs) in a four-dimensional NH system, finding a hypersphere-like non-Abelian Yang monopole in a five-dimensional parameter space, formed by EP2 pairs. Such an exotic structure enables the NH Yang monopole to exhibit a unique topological transition, which is inaccessible with the point-like counterpart. We characterize such a topological phenomenon with the second Chern number.
	\end{abstract}
	
	\maketitle
	Magnetic monopoles have long been focal points in the exploration of gauge theories and topological matters \cite{noauthor_quantised_1931}. 
    The Dirac monopole has been explored and constructed, and its deep physics related to Abelian gauge fields has been studied in a range of physical systems, including superconducting circuits \cite{roushan_observation_2014,PhysRevLett.113.050402,PhysRevLett.122.210401}, ultracold  atoms \cite{jotzu_experimental_2014,PhysRevLett.127.136802}, etc. However, the investigations of higher-order monopoles, such as the tensor monopole \cite {PhysRevD.9.2273,PhysRevLett.126.017702,chen_synthetic_2022} or the non-Abelian Yang monopole \cite{doi:10.1126/science.aam9031}, are relatively nascent. The Yang monopole is the key element of the Yang-Mills theory \cite{PhysRev.96.191,zee_quantum_2010}, representing a non-Abelian extension of the Dirac monopole, which necessitates the application of higher-order Chern numbers for its topological characterization \cite{PhysRevB.100.075423,sugawa_second_2018,PhysRevLett.117.015301}. Remarkedly, the Wilczek-Zee (WZ) phase theory \cite{PhysRevLett.52.2111} extends the concept of the Berry phase to degenerate subspaces, providing a powerful framework for characterizing non-Abelian operator-valued geometric phases, which is closely related to higher-order Chern numbers.
    The non-Abelian gauge field associated with Yang monopole is related to a gauge symmetry beyond that of quantum electrodynamics \cite{PhysRev.155.1554}, and serves as a foundational pillar in the standard model of particle physics.
    Over the two past decades, extensive interests have been stimulated in both theoretical and experimental exploration in synthetic non-Abelian gauge fields \cite{li_bloch_2016,di_liberto_non-abelian_2020,PhysRevLett.123.173202,PhysRevLett.95.010404,PhysRevLett.95.010403,PhysRevA.79.023624,PhysRevLett.102.080403,bermudez_topological_2010,PhysRevLett.109.145301,phuc_controlling_2015,leroux_non-abelian_2018,yang_synthesis_2019,sugawa_wilson_2021,PhysRevLett.58.2281,PhysRevA.42.3107,Zheng_2022}, as well as in related applications in holonomic quantum computation \cite{PhysRevLett.116.140502,Zhang2024,han2020experimentalrealizationuniversaltimeoptimal,PhysRevA.87.052307,duan_geometric_2001,abdumalikov_jr_experimental_2013,zu_experimental_2014,arroyo-camejo_room_2014,xu_demonstration_2021} and quantum signal processing \cite{singh2025nonabelianquantumsignalprocessing}.
	
	To date, the majority of investigations of high-order monopoles have been focused on closed systems \cite{weisbrich_tensor_2021,PRXQuantum.2.010310}. However, any system is subjected to environmentally-induced dissipation. For the no-jump case, the system dynamics is described by a non-Hermitian (NH) conditional Hamiltonian \cite{moiseyev_non-hermitian_2011,rotter_non-hermitian_2009,berry_physics_2004,heiss_physics_2012}. Due to the non-Hermiticity, both the eigenvalues and eigenstates coalesce at exceptional points (EPs), leading to many novel features, such as spectral real-to-complex transitions \cite{PhysRevLett.86.787,PhysRevLett.104.153601,gao_observation_2015,zhang_observation_2017}, chiral behaviors \cite{PhysRevX.8.021066,doppler_dynamically_2016,xu_topological_2016,yoon_time-asymmetric_2018,PhysRevLett.126.170506,PhysRevLett.124.070402,ren_chiral_2022}, reversed pump dependence of lasers \cite{PhysRevLett.108.173901}, parity–time symmetry \cite{PhysRevLett.80.5243,PhysRevLett.89.270401,ozdemir_paritytime_2019,PhysRevLett.103.093902,feng_single-mode_2014,hodaei_parity-timesymmetric_2014}, topology features \cite{bergholtz_exceptional_2021,ding_non-hermitian_2022}, exceptional entanglement transition \cite{PhysRevLett.131.260201} and criticality-enhanced sensitivity \cite{miri_exceptional_2019,chen_exceptional_2017,hodaei_enhanced_2017}. 
    The NH extension of the monopoles and the related exceptional topology have been extensively studied in both classical \cite{PhysRevLett.118.045701,zhen_spawning_2015,cerjan_experimental_2019,PhysRevLett.129.084301,tang_realization_2023} and quantum systems \cite{han_measuring_2024,ZHANG20252446}, but limited to the Abelian case associated with EP2s or EP3s \cite{yang2025exceptionalsurfacetopology,wu_third-order_2024,Chen2025Quantum,Zhang2025}. 
    In this work, we investigate the exceptional structure of a four-dimensional (4D) open system, whose quantum state evolution trajectory without quantum jumps is governed by a NH Hamiltonian featuring the competition between coherent couplings and dissipation. We find that there exists a structured Yang monopole, which corresponds to an exceptional hypersphere (EHS) consisting of degenerate EP2 pairs. The involved topology is characterized by the second Chern number, which is obtained by the integration of the non-Abelian curvature over the 5D parameter manifold \cite{manton_topological_2004}. We find that the Chern number makes a transition from 0 to 1 when the parameter-space manifold, initially being inside the  monopole, is progressively enlarged so as to enclose the monopole. Also, the related geometry can be described by the WZ geometric phase obtained by integrating the non-Abelian Berry connection along a closed loop.

    \textit{The non-Abelian exceptional hypersphere.}\textemdash We consider a four-level system where a synthetic non-Abelian gauge field can be established. The Hamiltonian is written as $H(q)=\vec{q}\cdot \vec{\Gamma}$, where $\vec{\Gamma}=\{\Gamma_1,\Gamma_2,\Gamma_3,\Gamma_4,\Gamma_5\}$ are a group of $4\times4$ Dirac matrices satisfying the Clifford algebra $\{\Gamma_i,\Gamma_j\}=2\delta_{ij}I_0^{4*4}$ \cite{PhysRev.96.191}, where $\{\cdot\}$ denotes the anti-commutator, $\delta_{ij}$ is the Kronecker delta and $I_0^{4*4}$ is the $4\times4$ identity matrix. In the presence of an NH term $i\kappa{\Gamma_4}$ $(\kappa>0)$ associated with population gain and loss, the Hamiltonian is transformed into 
	\begin{align}
		H(q)=\vec{q}\cdot \vec{\Gamma}+i\kappa{\Gamma_4}.
		\label{eq01}
	\end{align}%
	\begin{figure}[h!] 
		\centering
		\includegraphics[width=3.4in]{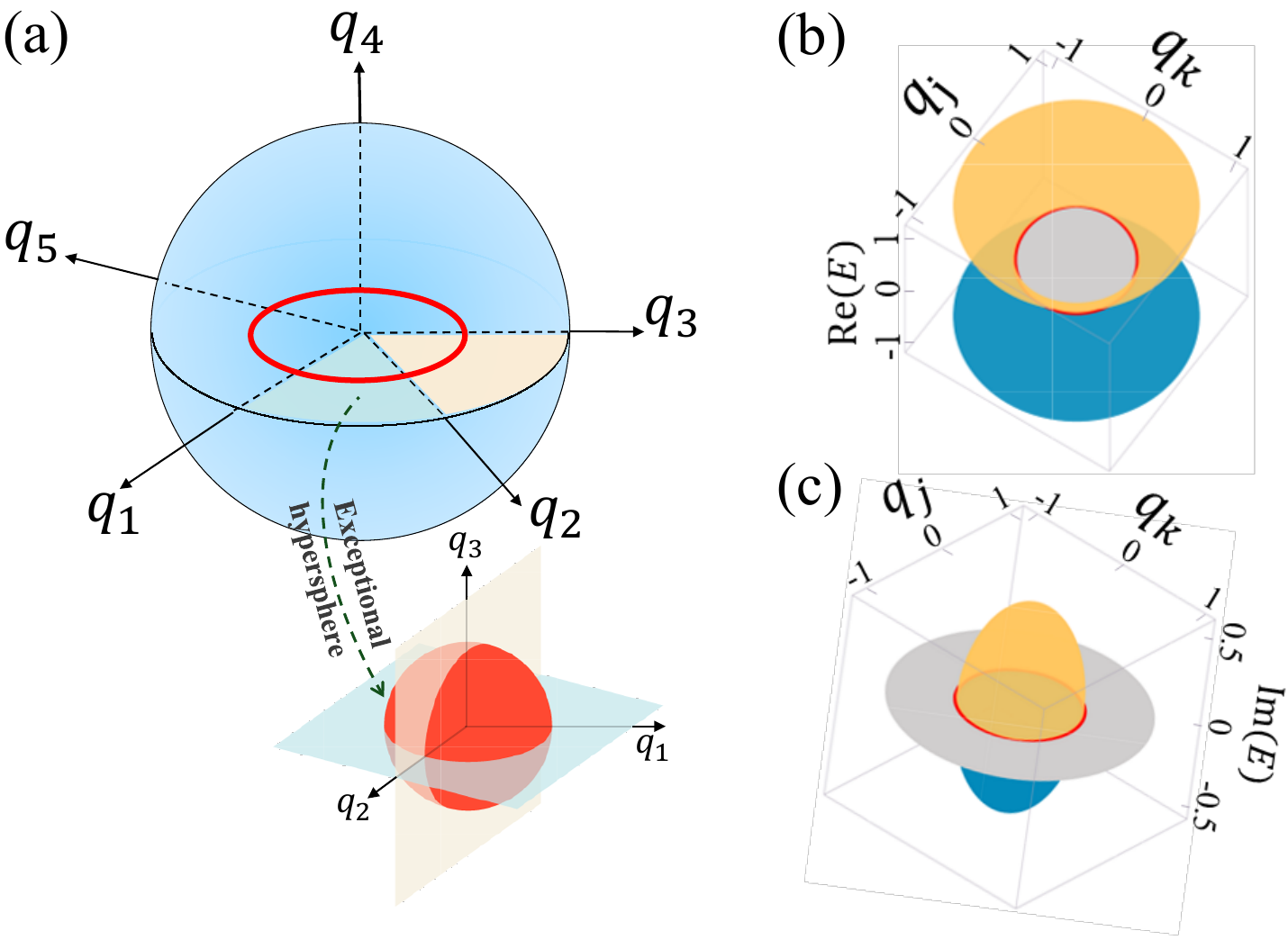}
		\caption{Conceptual schematic and energy spectra. (a) Schematic of the EHS in 5D parameter manifold, depicted as a hypersphere (red ring). Projections onto any 3D subspace manifested as spherical shells (red surfaces). Real (b) and imaginary (c) parts of the spectra with respect to $q_{j,k}\in{\{q_1,q_2,q_3,q_5\}}$.}
		\label{nFig1}
	\end{figure}%
	For Eq. (\ref{eq01}), there exists two pairs of degenerate real eigenvalues $E=\pm|q|$ when $\kappa=0$, which can be regarded as Yang monopole locating at the point $\vec{q}=0$ in the parameter space. With the introduction of an NH term, the Yang monopole morphs into a three-dimensional exceptional hypersphere (EHS) on the $q_4=0$ plane characterized by $\sum{q_n}^2=\kappa^2$. In Fig.~\ref{nFig1}(a), the EHS is shown as two spheres of radius $\kappa$ in the projections 3D parameter spaces $\{{q_1},{q_2},{q_3}\} ({q_5}=0)$ and $\{{q_1},{q_3},{q_5}\} ({q_2}=0)$. On this EHS, both real and imaginary components of the eigenvalues vanish, as shown in Fig.~\ref{nFig1}(b) and \ref{nFig1}(c), while two pairs of degenerate eigenstates coalesce into a single pair. Notice that the two degenerate eigenvalues satisfy $E_\pm=\pm\sqrt{|q|^2-\kappa^2}$, are purely real outside the EHS, while purely imaginary within the EHS [Fig.~\ref{nFig1}(b) and \ref{nFig1}(c)].
    
\begin{figure}[t] 
		\centering
		\includegraphics[width=3.4in]{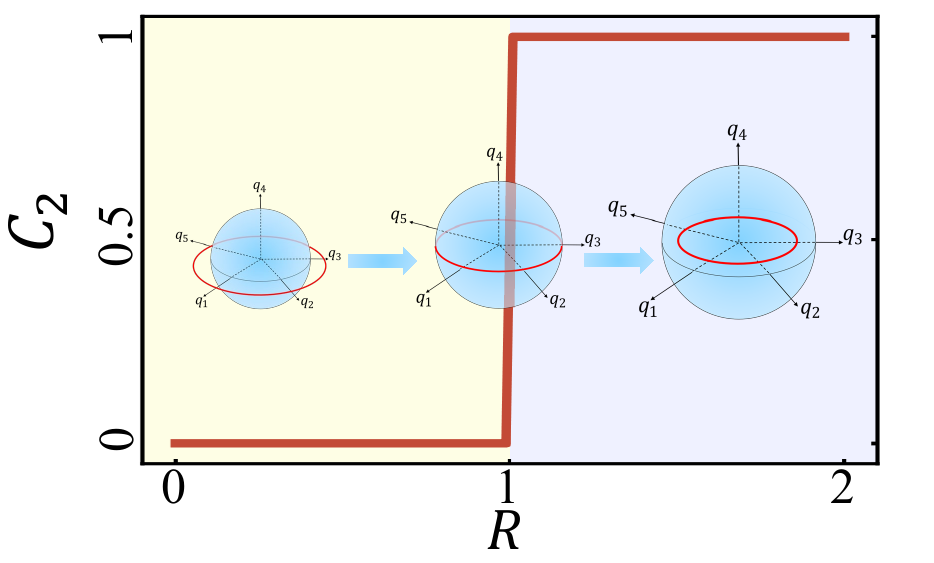}
		\caption{The second Chern number ($C_2$) versus the radius R of 5D parameter manifold. $C_2$ reaches 1 when the parameter manifold (projected onto blue sphere) encloses the EHS (projected onto red ring), and changes to 0 when unenclosed, with a sharp transition occurring at the critical boundary of $R=1$.}
		\label{nFig2}
	\end{figure}

    The non-Abelian EHS as Yang monopole can be characterized by the second Chern number ($C_2$), which is defined according to the $n$-wedge product of the non-Abelian curvature \cite{manton_topological_2004}
	\begin{eqnarray}
		C_n=\frac{1}{\left(2\pi \right)^n}\int{\frac{1}{n!}}\bigwedge F_{j,k} \textrm{d}^{2n}k,
        \label{eq02}
	\end{eqnarray}
	where $F_{jk}$ is the non-Abelian curvature whose element $F_{jk}^{\alpha\beta}$ is defined as $F_{jk}^{\alpha\beta}=\partial_j A_k^{\alpha\beta}-\partial_k A_j^{\alpha\beta}+i\left[A_j,A_k \right]^{\alpha\beta}$ \cite{PhysRev.56.340,singh_extensions_1989}. $A_j^{\alpha\beta}$ is the Berry connection in the form of a $2\times2$ matrix
	\begin{eqnarray}
		A^{\alpha\beta}_j=-i
		\begin{pmatrix}
			\langle\tilde{\psi}_-^{\alpha}|\partial_j\psi_-^{\alpha}\rangle &  \langle\tilde{\psi}_-^{\alpha}|\partial_j\psi_-^{\beta}\rangle \\
			\langle\tilde{\psi}_-^{\beta}|\partial_j\psi_-^{\alpha}\rangle & \langle\tilde{\psi}_-^{\beta}|\partial_j\psi_-^{\beta}\rangle \\
		\end{pmatrix},
        \label{eq03}
	\end{eqnarray}
    where $|\psi_-^{\alpha,\beta}\rangle$ are two eigenstates corresponding to the lower eivenvalues and $\langle\tilde{\psi}_-^{\alpha,\beta}|$ are the normalized left eigenstate of $|\psi_-^{\alpha,\beta}\rangle$. Without loss of generality, we set $\kappa=1$ and assume the parameter of $\vec{q}=R\{\textrm{sin}\theta_1\textrm{sin}\theta_2\textrm{cos}\phi_2,\textrm{sin}\theta_1\textrm{cos}\theta_2\textrm{cos}\phi_1,\textrm{sin}\theta_1\textrm{cos}\theta_2\textrm{sin}\phi_1,
	\textrm{cos}\theta_1,\\ \textrm{sin}\theta_1
    \textrm{sin}\theta_2\textrm{sin}\phi_2\}$ to construct a manifold with the four angular coordinates $\{\theta_1, \theta_2, \phi_1, \phi_2\}$ and the radial coordinate $R$. $C_2$ is calculated according to Eqs. (\ref{eq02}) and (\ref{eq03}) (see Supplemental Material), whose values with respect to $R$ is shown in Fig.~\ref{nFig2}. $C_2$ is equal to 1 or 0, dependent on whether the 5D parameter manifold encloses ($R>1$) or does not enclose ($R<1$) the EHS, with the critical boundary at $R=1$ where the $C_2$ jumps between the two values, revealing topological transition inherent in such a 4D NH model. Note that the topological charge here is entirely carried by the whole EHS.     
		\begin{figure*}[t!] 
		\centering
		\includegraphics[width=6.8in]{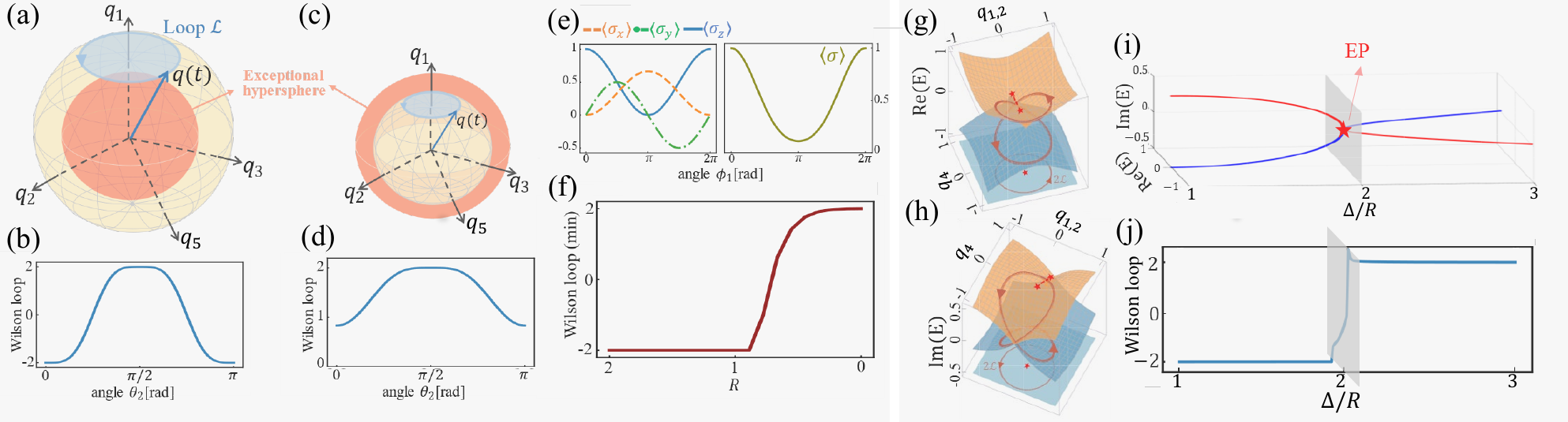}
		\caption{The characterization of the Wilson loop and of its relations. (a) Parameter manifold (projected onto yellow sphere) enclosing the EHS (projected onto orange sphere). The traversing path $\mathcal{L}$ to acquire (b) the Wilson loop $W_\mathcal{L}$ at distinct phase angle $\theta_2$. (c) Trajectory of loop $\mathcal{L}$ when the parameter manifold is within the EHS and (d) its corresponding $W_\mathcal{L}$ versus $\theta_2$. Expectation values of the three Pauli operators $\langle \sigma_j\rangle$ (left) and of their square summation $\langle\sigma^2\rangle$ (right) under the system evolution with respect to angle $\phi_1$ (e) for the traversing path $\mathcal{L}$ with $\theta_2=\pi/4$. (f) Minimal values of $W_\mathcal{L}$ versus the radius $R$ of the parameter manifold. (g) Real and (h) imaginary parts of the spectra on the Riemann surface for path $\mathcal{L}$ encircling the ring which is the projection of the EHS in $\{q_{1},q_2,q_4\}$ subspace. (i) Both real and imaginary parts of the spectra, where the EP is the projection of the EHS in $q_{1,2}$ axis, and (j) the corresponding $W_\mathcal{L}$ as function of $\Delta/R$, respectively, when path $\mathcal{L}$ encircling ($\Delta/R<2$) and non-encircling ($\Delta/R>2$) the EHS. The shadow region marks the critical boundary at $\Delta/R = 2$ where the transition happens.}
		\label{nFig3}
	\end{figure*}
	
	From other point of view, we can also employ the WZ geometric phase to characterize the exceptional physics of such an EHS. The unitary evolution operator characterizing the system dynamics can be described by the WZ phase factor \cite{PhysRevLett.52.2111}
	\begin{eqnarray}
		U_{\mathcal{L}}=\mathcal{P}\textrm{exp}\left(i\int_{\mathcal{L}}A_{j}^{\alpha\beta}\textrm{d}{j}\right),
        \label{eq04}
	\end{eqnarray}
    where $\mathcal{P}$ denotes the path-ordered integral along the path $\mathcal{L}$. The Wilson loop can be obtained by tracing out the $U_{\mathcal{L}}$: $W_\mathcal{L}=\textrm{tr}(U_{\mathcal{L}})$. For the model in Eq.~(\ref{eq01}), we parameterize the space as $\vec{q}=R\{\textrm{sin}\theta_2,\textrm{cos}\theta_2\textrm{cos}\phi_1,\textrm{cos}\theta_2\textrm{sin}\phi_1,0,0 \}$ ($\phi_1\in[0,2\pi]$). When $R>1$, the closed loop $\mathcal{L}$ traced by the control vector $\vec{q}$ is depicted in Fig.~\ref{nFig3}(a). Fig.~\ref{nFig3}(b) shows the Wilson loop, defined as $W_\mathcal{L}=\textrm{tr}\left[\mathcal{P}\textrm{exp}\left(i\int_{0}^{2\pi}A_{\phi_1}\textrm{d}{\phi_1}\right)\right]$, versus $\theta_2 (\in[0,\pi])$. At $\theta_2=0$, $\int A^{11}=-\int A^{22}=\pi$, which corresponds to the global minimum $W_{\mathcal{L}}=-2$; while at $\theta_2=\pi/2$, $\int A^{11}=\int A^{22}=0$, corresponding to the global maximum $W_{\mathcal{L}}=2$ (see Supplemental Material). When $R<1$, the parameter manifold does not enclose the EHS, the trajectory of the closed loop ${\mathcal{L}}$ and the corresponding $W_{\mathcal{L}}$ are depicted in Figs.~\ref{nFig3}(c) and \ref{nFig3}(d), respectively. Fig.~\ref{nFig3}(e) (left) displays the evolution of the expectation values of Pauli operators $\langle\sigma_j\rangle$ ($j=x,y,z$) within the degenerate ground subspace as the parameter $\phi_1$ traverses a closed loop from 0 to $2\pi$, with the system starting from one of the degenerate ground states at $\theta_2 = \pi/4$. This leads to their square summation $\langle\sigma^2\rangle\equiv\langle\sum\sigma_j^2\rangle<1$, as shown in Fig.~\ref{nFig3}(e) (right). This demonstrates the non-Abelian feature of our structured 4D NH system, in stark contrast with its Abelian counterpart of $\langle\sigma^2\rangle=1$.  
	Fig.~\ref{nFig3}(f) shows the minimal $W_{\mathcal{L}}$ versus the radius $R$ of the parameter manifold over $\theta_2\in[0,\pi]$. Remarkably, 
    it is found that, the minimal $W_{\mathcal{L}}$ remains $-2$ and keeps stable when the parameter manifold encloses the EHS; while it goes through a sharp increase when the parameter manifold crosses the EHS, and then smoothly increases, tending to $2$. It should be noticed that the Wilson loop $W_{\mathcal{L}}$ is not a topological quantity, and thus a sharp change in its values does not mean that the system undergoes a topological transition, which only represents a manifestation of the inherent geometric feature.    
    
	A further intriguing aspect arises when the evolution loop encircles the ring (projection of the EHS in $\{q_{1},q_2\}$ subspace), which gives arise to a remarkable spectral structure, and leads to exotic behaviors in the corresponding WZ geometric phase. We assume the parameter loop as $\vec{q}=\{(R\sin{\theta_1}+\Delta)/\sqrt{2},(R\sin{\theta_1}+\Delta)/\sqrt{2},0,R\cos\theta_1,0\}$, which corresponds to a loop of the radius $R$
	on $\{q_{1,2},q_4\}$ plane, with $\Delta$ denoting its distance from the origin of the loop and $q_{1,2}$ being the parameter direction of $\vec{q_1}+\vec{q_2}$. Figs.~\ref{nFig3}(g) and \ref{nFig3}(h) show the Riemann surface formed by the two degenerate eigenvalues when the parameter loop encircles the projected ring. After two full cycles of parameter evolution ($\theta_1$ mod 4$\pi$), the Wilson loop is $W_{2\mathcal{L}}=-2$. This is in contrast with the Hermitian case where the same value of $W_{1\mathcal{L}}$ is obtained through only one cycle of the parameter evolution. This distinctive difference is essentially due to the Möbius structure of the NH eigenspectrum, which indicates that it requires two full evolution cycles for the eigenenergies to return to their original values. 
    
    The intriguing geometric behaviors can be further revealed by changing $\Delta$, which displaces the parameter loop with respect to the EHS ring. Fig.~\ref{nFig3}(i) shows the transition of two degenerate eigenvalues from imaginary to real upon the parameter loop traversing the EP (projection of the EHS in $q_{1,2}$ axis). Correspondingly, when the parameter loop crosses the EP, the Wilson loop $W_{2\mathcal{L}}$ changes abruptly from -2 to 2, as shown in Fig.~\ref{nFig3}(j), which reflects the imaginary-to-real spectra transition in the NH model. 	
		\textit{Realization in a dissipative circuit QED system.}\textemdash For the experimental implementation of the 4D NH system, we consider a circuit QED model where two superconducting qutrits ($Q_1$, $Q_2$), whose three states are denoted as $|g\rangle$, $|e\rangle$ and $|f\rangle$, coupled to a resonator ($R$) which stores a quantized dissipative mode. The $|g\rangle \leftrightarrow |e\rangle$ transition is resonant with the resonator mode. The Hamiltonian is given by ($\hbar=1$ is set)
        \begin{eqnarray}
            H &=& \sum_n^{1,2}\left(\omega_{e_n}|e\rangle_n\langle e|+\omega_{f_n}|f\rangle_n\langle f|\right)+\omega_ra^\dagger a\nonumber\\
            &&+\left[\sum_n^{1,2}g_r\left(|g\rangle_n\langle e| a^\dagger+\sqrt{2}|e\rangle_n\langle f|a^\dagger\right)+H.c.\right],
        \end{eqnarray}
    \begin{figure}[h!] 
        \centering
        \includegraphics[width=3.4in]{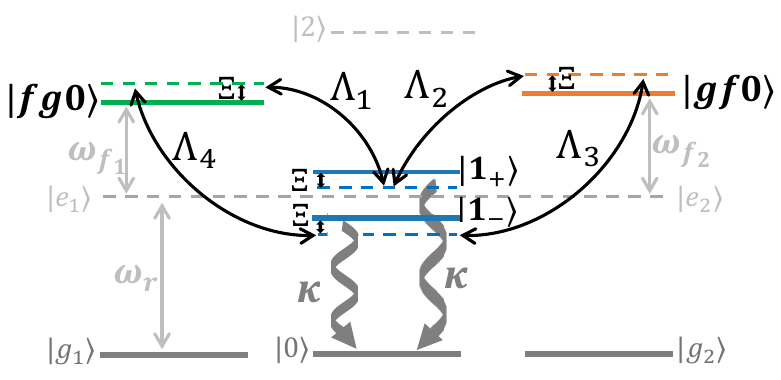}
        \caption{Schematic of the experimental protocol. Schematic of the driving strategy with circular coupling: $\{|fg0\rangle\xleftrightarrow{\Lambda_1} |1_+\rangle\xleftrightarrow{\Lambda_2}|gf0\rangle\xleftrightarrow{\Lambda_3}|1_-\rangle\xleftrightarrow{\Lambda_4}|fg0\rangle\}$, where $\Lambda_j$ ($j=1,2,3,4$) are the effective coupling strengths between states, $\Xi$ describes the detuning, and $\kappa$ is the single-photon loss rate of the resonator mode.}
        \label{nFig4}
	\end{figure}%
        where $\omega_{e_n}/2\pi$ ($\omega_{f_n}/2\pi$) and $\omega_{r}/2\pi$ denote frequencies of the $|e\rangle$- ($|f\rangle$)-state of $Q_{1,2}$ and $R$, respectively, $g_r$ is the $Q_n$-$R$ coupling strength and $a^\dagger$ ($a$) is the creation (annihilation) operator of the resonator. In order to obtain the required NH model of Eq.~(\ref{eq01}), we need four additional drives applied to the qutrits. The driving is described by the Hamiltonian $H_d=\sum_m^{1,2}\lambda_me^{i\left(\xi_mt+\phi_m\right)}\left(|g\rangle_1\langle e|+\sqrt{2}|e\rangle_1\langle f|\right)+\sum_m^{3,4}\lambda_me^{i\left(\xi_mt+\phi_m\right)}\left(|g\rangle_2\langle e|+\sqrt{2}|e\rangle_2\langle f|\right)+H.c.$, where $\lambda_m$, $\xi_m$ and $\phi_m$ are the amplitudes, frequencies and phases of the drives. With the choices $\xi_{1,2}=\omega_{f_1}-\omega_e+2\Xi\pm g_r/\sqrt{2}$ and $\xi_{3,4}=\omega_{f_2}-\omega_e+2\Xi\mp g_r/\sqrt{2}$, the four drives couples $|gf0\rangle$ and $|fg0\rangle$ to the dressed states $|1_{\pm}\rangle$ with the detuning $2\Xi$, respectively, where $|1_\pm\rangle=\left(|ge0\rangle/2+|eg0\rangle/2\pm|gg1\rangle/\sqrt{2}\right)$. The coupling induced by such drives essentially forms a circular structure, as shown in Fig.~\ref{nFig4}. Under the condition $\lambda_{m}\ll g_{r}$, the off-resonantly transitions can be discarded due to large detuning. In the interaction picture, the coherent dynamics is described by the effective Hamiltonian \cite{xu_demonstration_2021}
		\begin{eqnarray}
			H_I&=&\Xi\left(|fg0\rangle\langle fg0|+|gf0\rangle\langle gf0|-|1_+\rangle\langle1_+|-|1_-\rangle\langle1_-|\right)\nonumber\\
            &&+[\Lambda_1e^{i\phi_1}\left(|fg0\rangle\langle1_+|+|1_-\rangle\langle gf0|\right)\nonumber\\
			&&+\Lambda_2e^{i\phi_2}\left(|1_+\rangle\langle gf0|-|fg0\rangle\langle1_-|\right)]+H.c.,
		\end{eqnarray}
        where $\Lambda_{1,3}=\lambda_{1,3}/\sqrt{2}$, $\Lambda_{2}=-\lambda_2/\sqrt{2}$, $\Lambda_4=\lambda_4/\sqrt{2}$, $\phi_1=\phi_4$ and $\phi_2=\phi_3$. Considering the energy relaxation of the resonator $R$, the system dynamics is governed by the Lindblad master equation $d\rho(t)/dt=-i[H_{NH},\rho(t)]+\kappa a\rho a^\dagger$, where $\kappa$ is the single-photon loss rate of $R$, and the energy relaxation and dephasing of $Q$, and the dephasing of R have been ignored \cite{han_measuring_2024,song_continuous-variable_2017,PhysRevLett.123.060502,cpl_40_6_060301}. For the no-jump case, the system dynamics is governed by the NH Hamiltonian
		\begin{eqnarray}
			H_{NH}=H_I-\frac{1}{2}i\kappa a^\dagger a.
		\end{eqnarray}
        In the subspace $\mathcal S \in \{|gf0\rangle,|fg0\rangle,|1_{\pm}\rangle\}$, this NH Hamiltonian corresponds to Eq.~(\ref{eq01}).

        In order to characterize the second Chern number ($C_2$) and the Wilson loop ($W_{\mathcal{L}}$), the key is to extract the left and right degenerate eigenstates $|\psi_-^{\alpha,\beta}\rangle$ under different parameters for the manifold or loop, which in practice needs four steps \cite{PhysRevLett.131.260201,han_measuring_2024,ZHANG20252446}: (1) starting from an arbitrary initial state within $\mathcal S$, letting it evolve for specific times $t_j$; (2) at these $t_j$, mapping the resonator state to an ancilla qutrit $Q_a$, and measuring the three qutrits $Q_1$, $Q_2$, and $Q_a$ by joint quantum state tomography at three-state subspace $\{|g\rangle,|e\rangle,|f\rangle\}$ for each qutrit so as to extract the whole state information of the three qutrits, which corresponds to the state of $Q_1$, $Q_2$, and $R$ before state mapping; (3) postselecting the states within $\mathcal S$ by discarding the state $|gg0\rangle$ induced by quantum jumps and by renormalizing the whole state; (4) fitting the eigenstates $|\psi_-^{\alpha,\beta}\rangle$ through the least-quare method, and applying them in Eqs.~(\ref{eq02}) and (\ref{eq04}) to finally get $C_2$ and $W_\mathcal{L}$.

		
	\textit{Conclusion.}\textemdash In summary, we have shown that a point-like Yang monopole in a 5D parameter space is extended to a 3D hypersphere when a non-Hermitian term is introduced to the Hamiltonian of a 4D system. This NH monopole, formed by degenerate EP2 pairs, can display exotic topological transitions that are inaccessible with the Hermitian counterparts. We characterize the topological features by the second Chern number and the Wilson loop. We further proposed a scheme for realizing the dissipative 4D NH model in circuit QED.
		
        We thank Seiji Sugawa at University of Tokyo for valuable suggestions. This work was supported by the National Natural Science Foundation of China (Grant Nos. 12474356, 12475015, 12274080, 11875108).
		
		\bibliography{reference_C2}
	\end{document}


\title{Supplemental Material for "A hypersphere-like non-Abelian Yang monopole and its topological characterization"}
    \author{Shou-Bang Yang}
            \affiliation{Fujian Key Laboratory of Quantum Information and Quantum
        		Optics, College of Physics and Information Engineering, Fuzhou University,
        		Fuzhou, Fujian, 350108, China}
    \author{Pei-Rong Han} 
            \affiliation{Fujian Key Laboratory of Quantum Information and Quantum
    		Optics, College of Physics and Information Engineering, Fuzhou University,
    		Fuzhou, Fujian, 350108, China}
            \affiliation{School of Physics and Mechanical and Electrical Engineering, Longyan University, Longyan, China.}
	\author{Wen Ning}
            \affiliation{Fujian Key Laboratory of Quantum Information and Quantum
        		Optics, College of Physics and Information Engineering, Fuzhou University,
        		Fuzhou, Fujian, 350108, China}
	\author{Fan Wu}
            \affiliation{Fujian Key Laboratory of Quantum Information and Quantum
    		Optics, College of Physics and Information Engineering, Fuzhou    University,
    		Fuzhou, Fujian, 350108, China}
	\author{Zhen-Biao Yang} \email{zbyang@fzu.edu.cn}
            \affiliation{Fujian Key Laboratory of Quantum Information and Quantum
		Optics, College of Physics and Information Engineering, Fuzhou University,
		Fuzhou, Fujian, 350108, China}
	\author{Shi-Biao Zheng} \email{t96034@fzu.edu.cn}
            \affiliation{Fujian Key Laboratory of Quantum Information and Quantum
		Optics, College of Physics and Information Engineering, Fuzhou University,
		Fuzhou, Fujian, 350108, China}
	\vskip0.5cm
    \maketitle
	\tableofcontents
	
\section{The non-Abelian exceptional hypersphere}
\subsection{The second Chern number}
	In the Hermitian case, the Hamiltonian for a non-Abelian four-level system can be expressed as (set $\hbar=1$)
\begin{align}
		H &= \vec{q}\cdot \vec{\Gamma}=(q_1\Gamma_1+q_2\Gamma_2+q_3\Gamma_3+q_4\Gamma_4+q_5\Gamma_5)\nonumber\\
	 	&=
	 	\begin{pmatrix}
		 		q_4 & q_2+iq_3 & 0 & -q_1-iq_5 \\ 
		 		q_2-iq_3 & -q_4 & q_1+iq_5 & 0 \\
		 		0 & q_1-iq_5 & q_4 & q_2-iq_3 \\
		 		-q_1+iq_5 & 0 & q_2+iq_3 & -q_4\\
		 	\end{pmatrix}\nonumber\\
	 \nonumber\\
	 	&=
	 	\begin{pmatrix}
		 		\cos{\theta_1} & \sin{\theta_1}\cos{\theta_2}e^{i\phi_1} & 0 & -\sin{\theta_1}\sin{\theta_2}e^{i\phi_2} \\ 
		 		\sin{\theta_1}\cos{\theta_2}e^{-i\phi_1} & -\cos{\theta_1} & \sin{\theta_1}\sin{\theta_2}e^{i\phi_2} & 0 \\
		 		0 & \sin{\theta_1}\sin{\theta_2}e^{-i\phi_2} & \cos{\theta_1} & \sin{\theta_1}\cos{\theta_2}e^{-i\phi_1} \\
		 		-\sin{\theta_1}\sin{\theta_2}e^{-i\phi_2} & 0 & \sin{\theta_1}\cos{\theta_2}e^{i\phi_1} & -\cos{\theta_1}\\
		 	\end{pmatrix},
	\label{01}
\end{align}
where $\vec{q}\equiv(q_1,q_2,q_3,q_4,q_5)=(\sin{\theta_1}\sin{\theta_2}\cos{\phi_2},\sin{\theta_1}\cos{\theta_2}\cos{\phi_1},\sin{\theta_1}\cos{\theta_2}\sin{\phi_1},\cos{\theta_1},\sin{\theta_1}\sin{\theta_2}\sin{\phi_2})$ and $\vec{\Gamma}=\{\Gamma_1,\Gamma_2,\Gamma_3,\Gamma_4,\Gamma_5\}$  are a group of $4\times4$ Dirac matrices \cite{PhysRev.96.191}
	 \begin{align}
		 	\Gamma_1=
		 	\begin{pmatrix}
			 		0 & 0&0&-1\\
			 		0&0&1&0\\
			 		0&1&0&0\\
			 		-1&0&0&0
			 	\end{pmatrix},
		 	\Gamma_2=
		 	\begin{pmatrix}
			 		0 & 1&0&0\\
			 		1&0&0&0\\
			 		0&0&0&1\\
			 		0&0&1&0
			 	\end{pmatrix},
		 	\Gamma_3=
		 	\begin{pmatrix}
			 		0 & i&0&0\\
			 		-i&0&0&0\\
			 		0&0&0&-i\\
			 		0&0&i&0
			 	\end{pmatrix},
		 	\Gamma_4=
		 	\begin{pmatrix}
			 		1 & 0&0&0\\
			 		0&-1&0&0\\
			 		0&0&1&0\\
			 		0&0&0&-1
			 	\end{pmatrix},
            \Gamma_5=
		 	\begin{pmatrix}
			 		0 & 0&0&-i\\
			 		0&0&i&0\\
			 		0&-i&0&0\\
			 		i&0&0&0
			 	\end{pmatrix},
		 \end{align}
satisfying the Clifford algebra $\{\Gamma_i,\Gamma_j\}=2\delta_{ij}I_0^{4\times4}$, where $\delta_{ij}$ is the Kronecker delta and $I_0^{4\times4}$ is the $4\times4$ identity matrix. In this context, the system's topological defect is interpreted as the presence of a Yang monopole at the origin. The topological defect can be characterized by the second Chern number ($C_2$)
\begin{eqnarray}
	C_2=\frac{1}{32\pi^2 }\int{\varepsilon^{\mu\nu\lambda\xi}\textrm{Tr}(F_{\mu\nu}F_{\lambda\xi}) \textrm{d}\theta_1\textrm{d}\theta_2\textrm{d}\phi_1\textrm{d}\phi_2},
	\label{S3}
\end{eqnarray}
	where $\mu,\nu,\lambda,\xi\in\{\theta_1,\theta_2,\phi_1,\phi_2\}$, $\varepsilon^{\mu\nu\lambda\xi}$ is the Levi-Civita symbol described in the four dimension, and the sum runs over repeated indices and $F_{\mu\nu}$ denotes the non-Abelian Berry curvature, which can be derived from the formula
	\begin{eqnarray}
		F_{\mu\nu}=\partial_\mu A_{\nu}-\partial_\nu A_{\mu}-i[A_\mu,A_\nu],
	\end{eqnarray}
	where $A_j$ $(j = \mu, \nu)$ is non-Abelian Berry connection defined by a $2\times2$ matrix
	\begin{eqnarray}
		A_\mu=i\begin{pmatrix}
			\langle\psi_\pm^\alpha|\partial_\mu\psi_\pm^\alpha\rangle & \langle\psi_\pm^\alpha|\partial_\mu\psi_\pm^\beta\rangle\\
			\langle\psi_\pm^\beta|\partial_\mu\psi_\pm^\alpha\rangle & \langle\psi_\pm^\beta|\partial_\mu\psi_\pm^\beta\rangle
		\end{pmatrix},
	\end{eqnarray}
	and $[A_\mu,A_\nu]\neq0$, with $|\psi_\pm^\alpha\rangle$ and $|\psi_\pm^\beta\rangle$ representing the two degenerate eigenstates which correspond to high and low energy levels.
	
	In the non-Hermitian (NH) case which considers the population gain and loss, the Hamiltonian of this system is rewritten as 
	\begin{eqnarray}\label{eqs6}
		H_{NH} = \vec{q}\cdot \vec{\Gamma}+i\kappa{\Gamma_4}.
	\end{eqnarray}
	\begin{figure}[t!] 
		\centering
		\includegraphics[width=6.0in]{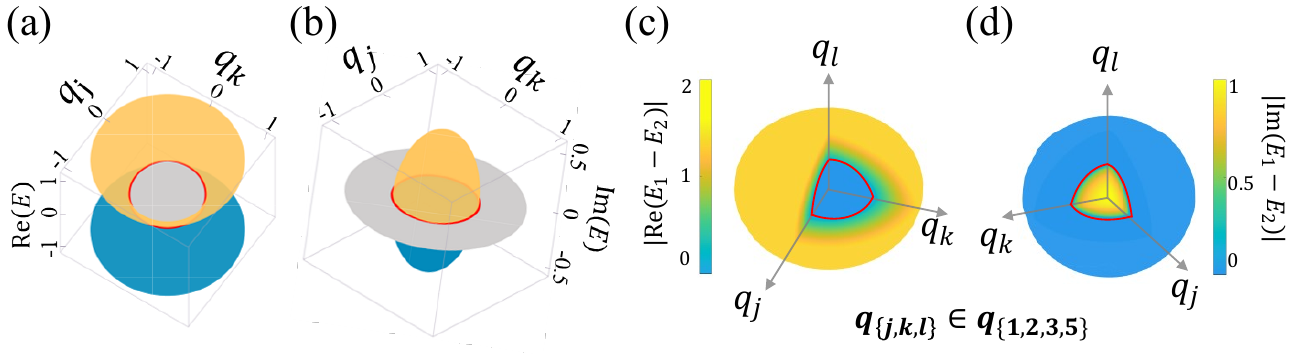}
		\caption{The EHS projected onto the lower-dimensional subspaces. The (a) real and (b) imaginary parts of the eigenvalues in an arbitrary 2D subspace $\{q_j,q_k\}\in\{q_1,q_2,q_3,q_5\}$, where the 2D projection of the EHS manifests as a ring (red ring) along which both the eigenenergies and eigenstates coalesce. Differences of the (c) real parts between two eigenvalues and those of the (d) imaginary parts, respectively, in the projected 3D subspace, where the EHS is projected as a spherical surface.}
		\label{sFig0}
	\end{figure}%
    The topological defect in the parameter space of Eq. (\ref{eqs6}) is extended from a single point to a hypersphere embedded in the four-dimensional parameter space spanned by $\{q_1,q_2,q_3,q_5\}$, which satisfies $q_4=0$ and $q_1^2+q_2^2+q_3^2+q_5^2=\kappa^2$. In contrast to the purely real degenerate eigenvalues in the Hermitian case, those in the NH system are purely real outside the exceptional hypersphere (EHS) and purely imaginary inside the EHS. The NH Hamiltonian in the five-dimensional (5D) parameter manifold can be structured as 
\begin{align}
	H_{NH} =
	\begin{pmatrix}
		R\cos{\theta_1}+i\kappa & R\sin{\theta_1}\cos{\theta_2}e^{i\phi_1} & 0 & -R\sin{\theta_1}\sin{\theta_2}e^{i\phi_2} \\ 
		R\sin{\theta_1}\cos{\theta_2}e^{-i\phi_1} & -R\cos{\theta_1}-i\kappa & R\sin{\theta_1}\sin{\theta_2}e^{i\phi_2} & 0 \\
		0 & R\sin{\theta_1}\sin{\theta_2}e^{-i\phi_2} & R\cos{\theta_1}+i\kappa & R\sin{\theta_1}\cos{\theta_2}e^{-i\phi_1} \\
		-R\sin{\theta_1}\sin{\theta_2}e^{-i\phi_2} & 0 & R\sin{\theta_1}\cos{\theta_2}e^{i\phi_1} & -R\cos{\theta_1}-i\kappa\\
	\end{pmatrix},
	\label{S7}
\end{align}
    spanned by the four bases within the subspace $\mathcal{S} \in \{|fg0\rangle,|1_+\rangle,|gf0\rangle,|1_-\rangle\}$. Figs.~\ref{sFig0}(a) and~\ref{sFig0}(b) display the real and imaginary parts of the eigenenergies in the two-demensional parameter subspace $\{q_j,q_k\}$, where the red ring is the projection of EHS. Figs.~\ref{sFig0}(c) and~\ref{sFig0}(d) show the eigenenergies in the three-dimensional subspaces $\{q_j,q_k,q_l\}$, in which case, the projection of the EHS is a spherical surface. For this NH Hamiltonian, the eigenvalues are $E_\pm=\pm\sqrt{R^2-\kappa^2+2i\kappa R\cos{\theta_1}}$, and the corresponding right eigenvectors are denoted as 
	\begin{eqnarray}
		|\psi_\pm^\alpha\rangle=
		\begin{pmatrix}
			R\sin{\theta_1}\\
			(E_\pm-R\cos{\theta_1}-i\kappa)\cos{\theta_2}e^{-i\phi_1}\\
			0\\
			-(E_\pm-R\cos{\theta_1}-i\kappa)\sin{\theta_2}e^{-i\phi_2}
		\end{pmatrix}/N_\pm,
		|\psi_\pm^\beta\rangle=
		\begin{pmatrix}
			0\\
			(E_\pm-R\cos{\theta_1}-i\kappa)\sin{\theta_2}e^{-i\phi_1}\\
			R\sin{\theta_1}\\
			(E_\pm-R\cos{\theta_1}-i\kappa)\cos{\theta_2}e^{-i\phi_2}
		\end{pmatrix}/N_\pm,
	\end{eqnarray}
	 with their adjoint left eigenvectors derived from the normalization condition $\langle\psi_m^j|\psi_n^k\rangle=\delta_{mn}\delta_{jk}$.
	
\begin{figure}[h!] 
	\centering
	\includegraphics[width=6.0in]{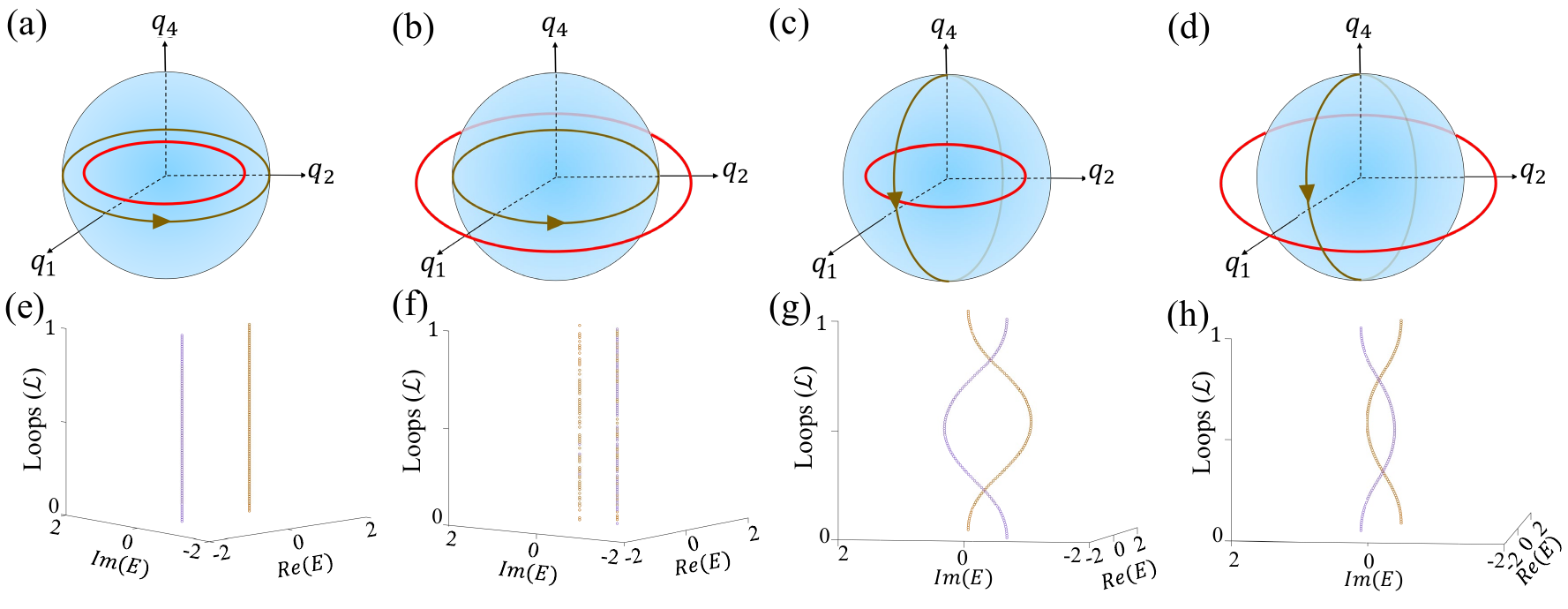}
	\caption{Schematic diagrams of the eigenvalues structures. (a) and (b) depict the rotation of parameters on $\{q_1,q_2\}$ plane over one full evolution ($\theta_2$ from 0 to $2\pi$) when the parameter hyperspherical respectively enclose and does not enclose the EHS, with $\theta_1$, $\phi_1$ and $\phi_2$ being zero. (c) and (d) demonstrate parameter rotation on $\{q_1,q_4\}$ plane as $\theta_1$ varies from 0 to $2\pi$ while maintaining $\theta_2$, $\phi_1$, and $\phi_2$ to be zero. (e)-(h) correspondingly display the variations in the real and imaginary parts of the system's eigenvalues during the parameter rotations.}
	\label{sFig1}
\end{figure}

	By integrating the trajectories of the four parameters as $\theta_1\in[0,\pi]$, $\theta_2\in[0,\pi/2]$, $\phi_1\in[0,2\pi]$ and $\phi_2\in[0,2\pi]$, we traverse the entire parameter hypersphere and subsequently compute $C_2$ to characterize the topological defect in this scenario as
	\begin{eqnarray}
		C_2=\frac{6R^7\sin^7{\theta_1}(E_+-E_-)^3[R^2-(E_++i)(E_-+i)]}{[N_+N_+^LN_-N_-^L(E_+-E_-)^2]^2}.
	\end{eqnarray}
    To reduce the topological transition, we continuously reduce the radius of the parameter hypersphere, to displace it from fully enclosing the EHS to the status where it no longer encloses the EHS. During this process, we observe the variation of $C_2$ from 1 to 0, which reveals the topological transition associated with the EHS, with the critical boundary where $C_2$ changes sharply between two values at $R = 1$. The transition of the eigenvalues versus $\theta_2$ and $\theta_1$ when the parameter space encloses the EHS or does not enclose the EHS are shown in Fig~\ref{sFig1}.

\subsection{The Wilczek-Zee phase and Wilson loop}
	The geometric phase for the non-Abelian case is characterized by the Wilczek-Zee (WZ) phase as
	\begin{eqnarray}
	U_{\mathcal{L}}=\mathcal{P}\textrm{exp}\left(i\int_{\mathcal{L}}A_\mu\textrm{d}\mu\right).
	\label{S10}
	\end{eqnarray}
\begin{figure}[t!] 
	\centering
	\includegraphics[width=6.0in]{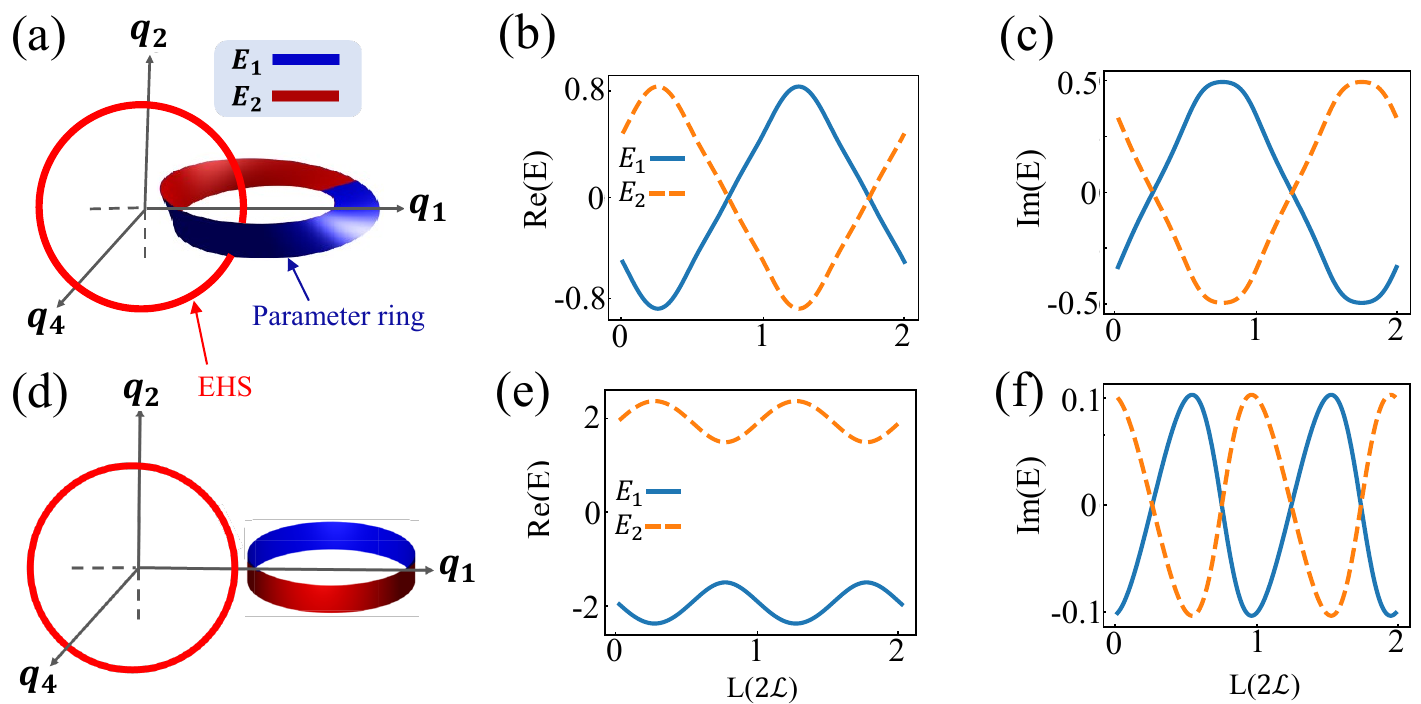}
	\caption{(a) When the parameter ring on $\{q_1,q_4\}$ plane encircles the EHS (projected as a ring on $\{q_1,q_2\}$ plane), the two degenerate eigenvalues exhibit a Möbius structure. The (b) real and (c) imaginary parts of the eigenvalues with the parameter ring traveling 2$\mathcal{L}$ path along the EHS. The blue and yellow lines represent the two pairs of degenerate eigenvalues $E_{1}$ and $E_{2}$, and both of which return to their original values after two circles along the EHS. (d)-(f) present the case when the parameter loop no longer encircles the EHS.}
	\label{sFig2}
\end{figure}
	We first consider a parameter space of Eq.~(\ref{S7}). For $R>1$, the closed loop $\mathcal{L}$ for the WZ phase surrounds the EHS; while for $R<1$, it resides within the EHS. Fixing $\theta_1=\pi/2$ and $\phi_2=0$, we vary $\phi_1\in[0,2\pi]$. The WZ phase accumulated by the evolution along $\mathcal{L}$ depends on the initial angle $\theta_2$. The system Hamiltonian in such a case is given by
	\begin{align}
		H_1 =
		\begin{pmatrix}
			i & R\cos{\theta_2}e^{i\phi_1} & 0 & -R\sin{\theta_2} \\ 
			R\cos{\theta_2}e^{-i\phi_1} & -i & R\sin{\theta_2} & 0 \\
			0 & R\sin{\theta_2} & i & R\cos{\theta_2}e^{-i\phi_1} \\
			-R\sin{\theta_2} & 0 & R\cos{\theta_2}e^{i\phi_1} & -i\\
		\end{pmatrix},
	\end{align}
	with its eigenenergies and eigenstates denoted as
	\begin{eqnarray}
		E_\pm=\pm\sqrt{R^2-1},
	\end{eqnarray}
	and 
	\begin{eqnarray}
		|\psi_\pm^\alpha\rangle=
		\begin{pmatrix}
			-(E_\pm+i)/R\\
			-e^{-i\phi_1}\cos{\theta_2}\\
			0\\
			\sin{\theta_2}
		\end{pmatrix}/N_\pm,
		|\psi_\pm^\beta\rangle=
		\begin{pmatrix}
			0\\
			\sin{\theta_2}\\
			(E_\pm+i)/R\\
			e^{i\phi_1}\cos{\theta_2}
		\end{pmatrix}/N_\pm,
	\end{eqnarray}
	respectively. The corresponding non-Abelian Berry connection is
	\begin{eqnarray}
		A_{\phi_1}=
		\begin{pmatrix}
			\cos^2{\theta_2}/N^2 & e^{i\phi_1}\cos{\theta_2}\sin{\theta_2}/N^2\\
			e^{-i\phi_1}\cos{\theta_2}\sin{\theta_2}/N^2 & -\cos^2{\theta_2}/N^2
		\end{pmatrix}.
	\end{eqnarray}
	A crucial note is that Eq.~(\ref{S10}) represents a path-ordered integral. Since $A_{\phi_1}$ does not commute between different points along the trace, a simple integration of its individual matrix elements is invalid. To compute the WZ phase factor $U_\mathcal{L}$, one must solve the differential equation
	\begin{eqnarray}
		\frac{\partial U_\mathcal{L}}{\partial\phi_1}=\partial_{\phi_1}\mathcal{P}\textrm{exp}\left(i\int_{\mathcal{L}}A_{\phi_1}\textrm{d}\phi_1\right)=iA_{\phi_1}U_\mathcal{L}
	\end{eqnarray}
	 with the initial condition $U_\mathcal{L}(\phi_1=0)=I$. When $\phi_1=2\pi$, this procedure yields the matrix elements as
	 \begin{eqnarray}
	 	U_\mathcal{L}^{11}&=&-\frac{\left(\sqrt{N^4-2N^2+2+2\cos{2\theta_2}-2N^2\cos{2\theta_2}}\right)\cos{\left(\sqrt{N^4-2N^2+2+2\cos{2\theta_2}-2N^2\cos{2\theta_2}}/N^2\right)}}{\sqrt{N^4-2N^2+2+2\cos{2\theta_2}-2N^2\cos{2\theta_2}}}\nonumber\\
	 	&&+\frac{i\left(2\cos^2{\theta_2}-N^2\right)\sin{\left(\sqrt{N^4-2N^2+2+2\cos{2\theta_2}-2N^2\cos{2\theta_2}}/N^2\right)}}{\sqrt{N^4-2N^2+2+2\cos{2\theta_2}-2N^2\cos{2\theta_2}}},                  \nonumber\\
	 	U_\mathcal{L}^{12}&=&U_\mathcal{L}^{21}=-\frac{i\sin{2\theta_2}\sin{\left(\sqrt{N^4-2N^2+2+2\cos{2\theta_2}-2N^2\cos{2\theta_2}}/N^2\right)}}{\sqrt{N^4-2N^2+2+2\cos{2\theta_2}-2N^2\cos{2\theta_2}}},               \nonumber\\
	 	U_\mathcal{L}^{22}&=&-\frac{\left(\sqrt{N^4-2N^2+2+2\cos{2\theta_2}-2N^2\cos{2\theta_2}}\right)\cos{\left(\sqrt{N^4-2N^2+2+2\cos{2\theta_2}-2N^2\cos{2\theta_2}}/N^2\right)}}{\sqrt{N^4-2N^2+2+2\cos{2\theta_2}-2N^2\cos{2\theta_2}}},\nonumber\\
	 	&&-\frac{i\left(2\cos^2{\theta_2}-N^2\right)\sin{\left(\sqrt{N^4-2N^2+2+2\cos{2\theta_2}-2N^2\cos{2\theta_2}}/N^2\right)}}{\sqrt{N^4-2N^2+2+2\cos{2\theta_2}-2N^2\cos{2\theta_2}}}. 
	 \end{eqnarray}
	The corresponding Wilson loop is $W_\mathcal{L}=\textrm{tr}(U_\mathcal{L})=U_\mathcal{L}^{11}+U_\mathcal{L}^{22}$.

	Remarkedly, the non-Abelian NH system exhibits another distinctive feature: a Möbius energy band structure. Since the EHS of the system can be projected as an exceptional ring (ER) on 
	$\{q_1,q_2\}$ plane within the $\{q_1,q_2,q_4\}$ subspace, and since the geometric WZ phase is defined over a closed loop $\mathcal{L}$—rather than a full spatial integral as in the second Chern number—when the path $\mathcal{L}$ is constrained to $\{q_{1,2},q_4\}$ plane, it can form an interwinding structure with the ER, as illustrated in Fig.~\ref{sFig2}(a). When $\theta_2=\pi/4$, $\phi_1=0$ and $\phi_2=0$, the system Hamiltonian is transformed into
	\begin{eqnarray}
		H_2=
		\begin{pmatrix}
			R\cos{\theta_1}+i&(R\sin{\theta_1}+\Delta)/\sqrt{2}&0& -(R\sin{\theta_1}+\Delta)/\sqrt{2}\\ 
			(R\sin{\theta_1}+\Delta)/\sqrt{2}&-R\cos{\theta_1}-i& (R\sin{\theta_1}+\Delta)/\sqrt{2}&0\\
			0&(R\sin{\theta_1}+\Delta)/\sqrt{2}&R\cos{\theta_1}+i& (R\sin{\theta_1}+\Delta)/\sqrt{2}\\
			-(R\sin{\theta_1}+\Delta)/\sqrt{2}&0&(R\sin{\theta_1}+\Delta)/\sqrt{2}& -R\cos{\theta_1}-i
		\end{pmatrix}.
		\label{S17}
		\end{eqnarray}
	Under this configuration, the energy bands exhibit a Möbius-like topology, meaning that the state returns to its original value only after traversing the closed path twice 2$\mathcal{L}$, rather than once as in the Hermitian case. The evolution of the real and imaginary parts of the eigenvalues along 2$\mathcal{L}$ is shown in Fig.~\ref{sFig2}(b) and (c), respectively. This results in a geometric WZ phase characterized by the Wilson loop $W_{2\mathcal{L}}=\textrm{tr}[\mathcal{P}\textrm{exp}\left(i\int_{0}^{4\pi}A_{\theta_1}\textrm{d}{\theta_1}\right)]$, which is calculate as $-2$. As $\Delta$ increases sufficiently enough such that the parameter loop no longer encloses the EHS, the WZ phase undergoes a sharp transition to $W_{2\mathcal{L}}=2$. Schematic illustrations of the corresponding energy band structure and the trajectories of the real and imaginary parts of eigenvalues are presented in Fig.~\ref{sFig2}(d)-(f).

	\bibliography{reference_sup}